\documentclass[conference]{IEEEtran}
\IEEEoverridecommandlockouts

\usepackage{cite}
\usepackage{amsmath,amssymb,amsfonts}
\usepackage{algorithm}
\usepackage{algorithmic}
\usepackage{diagbox}
\usepackage{subcaption}
\usepackage{adjustbox}
\usepackage{graphicx} 
\usepackage{subcaption}
\usepackage{stfloats}  
\usepackage{caption}    

\usepackage{footmisc}

\usepackage{multirow} 
\usepackage{pifont}
\usepackage{changes}
\definechangesauthor[name={Tongyu Song}, color=orange]{AF}

\usepackage{makecell}
\usepackage{graphicx}
\usepackage{hyperref}
\usepackage{textcomp}
\usepackage{xcolor}
\def\BibTeX{{\rm B\kern-.05em{\sc i\kern-.025em b}\kern-.08em
    T\kern-.1667em\lower.7ex\hbox{E}\kern-.125emX}}
\begin{document}

\title{D-LoRa: a Distributed Parameter Adaptation Scheme for LoRa Network
}
\author{\IEEEauthorblockN{Ruiqi Wang\textsuperscript{1}, Tongyu Song\textsuperscript{2*},
Jing Ren\textsuperscript{1*}, Xiong Wang\textsuperscript{1}, Shizhong Xu\textsuperscript{1}, Sheng Wang\textsuperscript{1}}
\IEEEauthorblockA{
\textsuperscript{1}\textit{University of Electronic Science and Technology of China, Chengdu, China}\\
\textsuperscript{2}\textit{Zhejiang Lab, Hangzhou, China}\\
202421011308@std.uestc.edu.cn, \\ 
tongyusong@ieee.org, \{renjing,wangxiong,xsz,wsh\_keylab\}@uestc.edu.cn
}
\thanks{* Corresponding Author}
}

\maketitle
 
\begin{abstract}
The deployment of LoRa networks necessitates joint performance optimization, including packet delivery rate, energy efficiency, and throughput, by dynamically configuring multiple LoRa parameters for packet transmission across varying channel environments. Because of the complex channel features modeling and the coupling relationship between LoRa parameters and metrics, existing works have opted to focus on certain parameters or specific metrics to simplify the problem, leading to limited adaptability. Therefore, we propose D-LoRa, a distributed parameter adaptation scheme, based on reinforcement learning towards network performance. We first build a comprehensive analytical model for the LoRa network that considers complex channel features, including path loss, quasiorthogonality of spreading factor, and packet collision. Then, we formulate the joint optimization problem as a combinatorial Multi-Armed Bandit (CMAB) problem and devise metric factors to handle the trade-off among different performance metrics. Experimental results show that our scheme can increase the packet delivery rate by up to 18.5\% and demonstrate superior adaptability across different performance metrics.
\end{abstract}

\begin{IEEEkeywords}
LoRa, CMAB, Packet Delivery Rate,
Energy Efficiency, Throughput, Distributed Parameter Adaptation
\end{IEEEkeywords}

\section{Introduction}
Long-Range (LoRa) is a prominent Low-Power Wide Area Network (LPWAN) communication technology widely used in smart cities, farm monitoring, and other Internet of Things (IoT) applications \cite{LoRaReview}. The performance of LoRa networks is primarily measured by three metrics, i.e., Packet Delivery Rate (PDR), Network Energy Efficiency (EE), and Throughput (TH). However, LoRa networks often suffer from performance degradation due to packet loss caused by collisions and propagation effects in time-varying communication channels \cite{QoC-A}. 
These problems can be effectively mitigated  by dynamically configured packet transmission parameters, including Spreading Factor (SF), Bandwidth (BW), Carrier Frequency (CF), and Transmission Power (TP) \cite{LoRaReview}. Therefore, implementing an effective parameter adaptation strategy for LoRaWAN  (the LoRa MAC layer protocol)  becomes imperative to minimize packet loss and enhance network performance.

However, optimizing packet transmission parameters in a LoRa system can be framed as a combinatorial optimization problem, which is NP-hard \cite{nphard1}. Moreover, enhancing LoRa network performance involves inherent trade-offs among these performance metrics due to the intricate coupling relationship between transmission parameters and performance metrics. Adjusting one parameter can simultaneously influence multiple metrics, often in conflicting ways. For instance, employing a smaller SF and larger BW reduces the packet's Time on Air (ToA), thereby enhancing network throughput. Yet, this configuration also results in a shorter transmission range, making packets from distant nodes more prone to loss and consequently reducing the network's PDR. Similarly, increasing TP can effectively improve the probability of successful packet reception, thus enhancing PDR, but it also leads to higher energy consumption, decreasing the network's EE. These coupling relationships underscore the complexity of parameter optimization, as simply improving one performance metric may adversely affect others. 

Recent studies have proposed various parameter adaptation schemes for LoRaWAN. Some works deploy centralized algorithms on a central entity (gateway or network server), configuring each node's parameters by gathering network-wide data and relaying configurations to the nodes via downlink messages \cite{LoRaReview}. Adaptive Data Rate (ADR) \cite{ADR},  a representative centralized algorithm, dynamically adjusts three parameters (SF, BW, TP). However, this kind of algorithm requires significant communication resources for node interactions, potentially causing congestion and latency issues \cite{LoRaReview}.

Other works focus on conducting distributed algorithms that allow nodes to self-assign parameters without global information. Hong et al. \cite{LR-RL} proposed LR-RL, a reinforcement learning (RL)-based method for SF allocation aimed at minimizing packet collision rates. Abdelghany et al. \cite{QoC-A} introduced QoC-A, which considers channel quality to select transmission channels and improve PDR. However, lacking global information, these algorithms often converge towards suboptimal solutions, and the performance gains remain constrained.

Researchers have recently modeled the distributed LoRa parameter adaptation problem as a Multi-Armed Bandit (MAB) problem, which can learn from historical transmission information to address the lack of global information. These approaches employ algorithms such as the Exponential weighting for Exploration and Exploitation (EXP3) \cite{LoRa-MAB} and the Upper Confidence Bound (UCB1) \cite{QoC-A}  to focus on optimizing single parameters (such as SF \cite{LoRa-MAB} or \cite{QoC-A}), demonstrating significant performance improvements in LoRa networks. However, due to the complex coupling between parameters and performance metrics, focusing solely on individual parameters or metrics may not yield the optimal overall network performance, while a joint optimization scheme for multiple parameters considering various performance metrics remains unexplored.

To fully explore the coupling relationships, this paper identifies a comprehensive analytical model. Then, D-LoRa, a distributed parameter adaptation algorithm based on Combinatorial Multi-Armed Bandit (CMAB) \cite{CMAB} is proposed, aiming to optimize three performance metrics (PDR, EE, TH) by allocating four LoRa parameters (SF, BW, CF, TP). Specifically, an agent is deployed at each node for decision-making regarding LoRa parameter adaptation. Configuring LoRa parameters is formulated as a CMAB, requiring each agent to manage four base arms, corresponding to four parameters. Furthermore, we devise metric factors for base arms to tailor the agent’s parameter preferences, dealing with the nonlinear trade-off relationships among the performance metrics to optimize diverse network metrics. The primary contributions of this paper are summarized as follows:
\begin{itemize}
\item 
\textbf{A comprehensive analytical model for LoRa networks:} We propose an analytical model for LoRa network considering the path loss, quasi-orthogonality of SF, and packet collision. We also formulate the problem of LoRa parameters adaption for packet transmission as an optimization problem based on this system model.
\item 
\textbf{Joint optimization scheme for LoRa multi-parameter adaptation problem:} We propose D-LoRa, which leverages CMAB learning for adapting LoRa packet transmission parameters (i.e., SF, BW, CF, and TP). The algorithm has been proved to be asymptotically optimal.
\item 
\textbf{Metric factors to handle the trade-off among different performance metrics:} Unique reward functions and metric factors have been devised for the base arms to optimize the metrics (i.e., PDR, EE, and TH) to meet the requirements of various application scenarios.
\item 
\textbf{Performance evaluation:} Simulations have been conducted using our simulator\cite{Simulator}. The experimental results show that our scheme can increase the PDR by up to 28.8\% compared to the best-performing baseline algorithm. Moreover, the diverse optimization results observed with D-LoRa's variant algorithms demonstrate the flexibility of D-LoRa.
\end{itemize}

\section{System Model and Problem Formulation}
\label{sec:model}
In this section, we first introduce the system modeling and then formulate the constrained optimization problem towards network performance improvement.

\subsection{Network Model}
In this paper, we consider a star topology network with a central gateway GW  and a set $\mathcal{N}=\{n_{1},n_{2},...,n_{N}\}$ of $N=|\mathcal{N}|$ nodes randomly distributed in the area. The nodes collect data from the surrounding environment and send the encoded packets to GW. The set $\mathcal{P}=\{p_{1},p_{2},...,p_{P}\}$ denotes the $P=|\mathcal{P}|$ packets sent by these nodes. The definitions are shown below.

\textit{Definition 1 (Gateway)}: The central gateway is denoted as $\text{GW} \doteq\left \langle x^{\text{g}},y^{\text{g}},\mathcal{P}^{\text{gr}}\right \rangle$, where $(x^{\text{g}},y^{\text{g}})$ is the two-dimensional coordinate of the gateway and $\mathcal{P}^{\text{gr}}$ is the set of the packets the gateway received.

\textit{Definition 2 (Node)}: A node $i$ is denoted as $n_{i}\doteq \left \langle x_{i},y_{i},\mathcal{P}^{\text{ns}}_{i},\mathcal{P}^{\text{nr}}_{i},\mathcal{P}^{\text{nl}}_{i},\mathcal{LR} \right \rangle$, where $(x_{i},y_{i})$ is the two-dimensional coordinate of the node. $\mathcal{P}^{\text{ns}}_{i}$ is the set of the packets sent by node $i$, $\mathcal{P}^{\text{nr}}_{i}$ is the set of the packets sent by node $i$ that are successfully received and $\mathcal{P}^{\text{nl}}_{i}$ is the set of the packets sent by node $i$ that are lost. $\mathcal{LR} \doteq \{\mathcal{SF},\mathcal{BW},\mathcal{CF},\mathcal{TP}\}$ is the set of available LoRa parameters for node $i$.

\textit{Definition 3 (Packet)}: A packet $j$ is denoted as $p_{j}\doteq\left \langle \text{PS}_{j},\text{id}(j),\mathcal{LP},E_{j}\right \rangle$, where $\text{PS}_{j}$ is the payload size of the packet and $\text{id}(j)$ is the identity of the node that sends packet $j$. 
$\mathcal{LP}=\{\text{SF}_{j},\text{BW}_{j},\text{CF}_{j},\text{TP}_{j}\}$ is the LoRa parameters selected by $n_{\text{id}(j)}$ to configure the packet. $E_{j}$ is the energy consumption of $n_{\text{id}(j)}$ sending $p_{j}$.

\subsection{Packet Collision Model}
\label{CollisionModel}
Authors in\cite{LoRaSim} proposed the definitions of the four conditions that can give rise to packet collisions: CF, SF, power, and timing. The binary variable $C_{j}$ indicating whether $p_{j}$ collides during transmission can be expressed as\cite{LoRaSim}:
\begin{equation}
   C_{j}=C^{\text{time}}_{j}\wedge C^{\text{SF}}_{j}\wedge C^{\text{CF}}_{j}\wedge C^{\text{pwr}}_{j},
    \label{Collision}
\end{equation}
where $C_{j}=1$ means that $p_{j}$ is lost due to collision during transmission.

\subsection{Packet Propagation Model}
Given the application scenarios in densely populated areas, we employ a radio transmission model based on the Log-Distance Path Loss Model. The path loss $L^{\text{pl}}_{j}(d)$ of $p_{j}$ when transmitted to the gateway can be expressed as:
\begin{equation}
   L^{\text{pl}}_{j}=\overline{L^{\text{pl}}}(d_{0})+10\gamma \text{log}(\frac{d}{d_{0}})+X_{\delta},
    \label{LoS}
\end{equation}
where $\overline{L^{\text{pl}}}(d_{0})$ is the average path loss with the reference distance is $d_{0}$, $d=\sqrt{(x_{\text{id}(j)}-x^{\text{g}})^{2}+(y_{\text{id}(j)}-y^{\text{g}})^{2}}$ is the Euclidean distance between $n_{\text{id}(j)}$ and GW, $\gamma$ is the path loss factor, $X_{\delta_{1}}\sim N(0,\delta_{1}^{2})$ is the normal distribution with mean $0$ and variance $\delta_{1}^{2}$ considering the shadowing effect. Assuming that the effects of other gains and losses in the propagation process are zero, the Received Signal Strength Indicator (RSSI) of $p_{j}$ at GW can be expressed as:
\begin{equation}
   \text{RSSI}_{j}=\text{TP}_{j}-\overline{L^{\text{pl}}}(d_{0})-10\gamma \text{log}(\frac{d}{d_{0}})-X_{\delta}.
    \label{RSSI}
\end{equation}
$p_{j}$ can only be successfully decoded by GW if $\text{RSSI}_{j}$ is not less than its Receiver Sensitivity ($\text{RS}$). $\text{RS}_{j}$ is determined by the $\text{\text{SF}}_{j}$ and $\text{BW}_{j}$ of $p_{j}$, which can be expressed as:
\begin{equation}
   \text{RS}_{j}=-174+10\log_{10}(\text{BW}_{j})+\text{NF}+\text{SNR}_{j},
    \label{RS}
\end{equation}
where $\text{NF}$ is the Noise Figure which is a fixed value depending on the hardware, and $\text{SNR}_{j}$ depends only on the $\text{\text{SF}}_{j}$. The $\text{RSs}$ corresponding to different SF and BW combinations are shown in TABLE \ref{tab1}\cite{Datasheet}:
\begin{table}[htbp]
\caption{RSs in dBm for different SF and BW combinations}
\centering
\begin{tabular}{c|c|c|c|c|c|c}
\hline
\diagbox{\textbf{BW}}{\textbf{\text{SF}}}&\textbf{7}&\textbf{8}&\textbf{9}&\textbf{10}&\textbf{11}&\textbf{12} \\
\hline
125kHz&-123&-126&-129&-132&-133&-136\\
\hline
250kHz&-120&-123&-125&-128&-130&-133\\
\hline
500kHz&-116&-119&-122&-125&-128&-130\\
\hline
\end{tabular}
\label{tab1}
\end{table}\\
Given the quasi-orthogonality of SF, we incorporated the inter-SF interference into the Signal-to-Interference-plus-Noise Ratio (SINR) calculations for packets. The $\text{SINR}_{j}$ of $p_{j}$ can be expressed as:
\begin{equation}
   \text{SINR}_{j}=\frac{\text{RSSI}_{j}}{\sum_{k\ne j,\text{\text{SF}}_{k}\ne \text{\text{SF}}_{j},\text{\text{CF}}_{k}= \text{\text{CF}}_{j}}\text{RSSI}_{k}+\text{N}_{0}\text{W(}u,\delta^{2})},
    \label{RS}
\end{equation}
where $\sum_{k\ne j,\text{\text{SF}}_{k}\ne \text{\text{SF}}_{j},\text{\text{CF}}_{k}= \text{\text{CF}}_{j}}\text{RSSI}_{k}$ is interference from the packets having reception overlap with $p_{j}$ with the same CF and different SF and $\text{N}_{0}\text{W}(u,\delta_{2}^{2})$ is the Additative White Gaussian Noise (AWGN). $p_{j}$ can only be successfully decoded by GW when $\text{SINR}_{j}$ is not less than its SINR threshold ($\text{SINR}^{\text{req}}_{j}$), which is determined by $\text{\text{SF}}_{j}$. The SINR thresholds for different SFs are shown in TABLE \ref{tab2}\cite{Datasheet}:
\begin{table}[htbp]
\caption{SINR thresholds in dB for different SF}
\centering
\begin{tabular}{c|c|c|c|c|c|c}
\hline
\textbf{\text{SF}}&\textbf{7}&\textbf{8}&\textbf{9}&\textbf{10}&\textbf{11}&\textbf{12} \\
\hline
$\textbf{\text{SINR}}^{\text{req}}$&-7.5&-10&-12.5&-15&-17.5&-20\\
\hline
\end{tabular}
\label{tab2}
\end{table}\\
Based on the analysis presented above, $p_{j}$ can only be successfully received by GW if both $\text{RSSI}_{j}$ and $\text{SINR}_{j}$ are not loss than the thresholds. $S_{j}$ is a binary variable that denotes whether $p_{j}$ is lost in the propagation process due to its weak signal strength, which can be expressed as:
\begin{equation}
   S_{j}=
   \begin{cases}
      0,& \text{if }\text{RSSI}_{j}\ge \text{RS}_{j}\text{ and }\text{SINR}_{j}\ge \text{SINR}^{\text{req}}_{j},\\
      1,& \text{otherwise. } 
    \end{cases}
    \label{LossFlag}
\end{equation}
\subsection{Energy Consumption Model}
The symbol period of $p_{j}$ is determined by $\text{\text{SF}}_{j}$ and $\text{BW}_{j}$, which is calculated as $T^{\text{sym}}_{j}=\frac{2^{\text{\text{SF}}_{j}}}{\text{BW}_{j}}$.  The Time on Air (ToA) of $p_{j}$ can be expressed as:
\begin{equation}
    \begin{aligned}
        \text{ToA}_{j}&=T^{\text{pre}}_{j}+T^{\text{pay}}_{j},
    \end{aligned}
    \label{ToA}
\end{equation}
where $T^{\text{pre}}_{j}$ is the preamble duration and $T^{\text{pay}}_{j}$ is the payload duration, which can be expressed respectively as:
\begin{equation}
        T^{\text{pre}}_{j}=(n^{\text{pre}}+4.25)\cdot T^{\text{sym}}_{j}\text{, } T^{\text{pay}}_{j}=n^{\text{pay}}_{j}\cdot T^{\text{sym}}_{j}, 
    \label{Duration}
\end{equation}
where $n^{\text{pre}}$ is the preamble size of a LoRa packet, which is 8 symbols in default. $n^{\text{pay}}_{j}$ is the number of payload symbols of packet $j$, which can be calculated as:
\begin{equation}
    \begin{aligned}
        n_{j}^{\text{pay}}&=(8+\max (\left \lceil \frac{8\text{PS}_{j}-4\text{\text{SF}}_{j}+28+16\text{CRC}-20\text{H}}{4(\text{\text{SF}}_{j}-2\text{DE})}\right \rceil\\& (\text{CR}+4),0)).
    \end{aligned}
    \label{n_payload}
\end{equation}
In the default configuration of LoRaWAN, Cyclic Redundancy Check (CRC) is enabled ($\text{CRC} = 1$), the header is enabled ($\text{H} = 0$), the LowDataRateOptimization is disabled ($\text{DE} = 0$), and the coding rate is set to $4/5$ ($\text{CR} = 1$). 
The energy consumed by $n_{\text{id}(j)}$ sending $p_{j}$ can be expressed as:
\begin{equation}
        E_{j}=\text{TP}_{j}\cdot \text{ToA}_{j}.
    \label{EnergyCon}
\end{equation}
\subsection{Optimal Problem Formulation}
In our work, we have considered $\text{PDR}$, $\text{EE}$, $\text{TH}$ as the network performance metrics. PDR is defined as the ratio of the number of packets successfully received by GW to the total number of packets sent by the nodes in the network, which can be expressed as:   
\begin{equation}
        \text{PDR}=\frac{\lvert \mathcal{P}^{\text{gr}} \rvert}{\sum_{i=0}^{N} \lvert \mathcal{P}^{\text{ns}}_{i} \rvert}.
    \label{PDR}
\end{equation}
EE is defined as the amount of effective data that can be successfully transmitted per unit of energy consumed by the network in bits/mJ, which can be expressed as:
\begin{equation}        
    \text{EE}=\frac{\sum_{p_{i}\in\mathcal{P}^{\text{gr}}}\text{PS}_{i}}{\sum_{n_{j}\in\mathcal N}\sum_{p_{k}\in \mathcal{P}^{\text{ns}}_{j}}E_{k}},
    \label{EE}
\end{equation}
where the numerator represents the size of effective data received by GW and the denominator represents the total energy consumption of the nodes during the transmission process. TH is defined as the amount of effective data successfully transmitted by the network per unit time in bps, which can be expressed as:
\begin{equation}
    \text{TH}=\frac{\sum_{p_{i}\in\mathcal{P}^{\text{gr}}}\text{PS}_{i}}{\sum_{n_{j}\in\mathcal N}\sum_{p_{k}\in \mathcal{P}^{\text{ns}}_{j}}\text{ToA}_{j}},
    \label{Throughput}
\end{equation}
where the denominator is the total time it takes for the nodes to send packets during the transmission process.

Our objective is to improve the performance of the LoRa network, which involves optimizing multiple metrics rather than focusing on a single one. 
Hence, we have devised a utility function, comprising a weighted aggregation of the three metrics, to assess the network's performance quantitatively. The performance optimization problem is formulated as following:
\begin{align}
    \text{(P1) }&\max \mathcal{U}=\theta \cdot \text{PDR} +\phi \cdot \text{EE} + \psi \cdot \text{TH}\label{problem} \\
    \text{s.t. } & C_{j}=0\land S_{j}=0, \forall p_{j}\in\mathcal{P}^{\text{gr}}\label{problemsub1}, \\
    & \text{SF}_{j}\in\mathcal{SF},\text{BW}_{j}\in\mathcal{BW}, \text{CF}_{j}\in\mathcal{CF}, \notag \\ 
    & \text{TP}_{j}\in\mathcal{TP}, \forall p_{j}\in\mathcal{P} \label{problemsub2}, 
\end{align}
where $\mathcal{U}$ is the utility function, $\theta$, $\phi$, $\psi$ are the weight factors and $\theta+\phi+\psi=1$. Constraint (\ref{problemsub1}) means $p_{j}$ can only be received by GW when both its $\text{RSSI}_{j}$ and $\text{SINR}_{j}$ are no less the thresholds and there is no collision occurred during its transmission. Constraints (\ref{problemsub2}) limit the LoRa parameters that are available for nodes to configure for packet tranmission.

\section{Distributed Parameter Adaptation Algorithm
}
\label{sec:method}
In this section, we solve the LoRa multi-parameter adaptation problem with CMAB-based method in a distributed manner. Then we present the design of rewards and the training process of our distributed adaptation algorithm D-LoRa.

\subsection{D-LoRa Design}
A naive centralized MAB approach involves a central entity (e.g., gateway) that maintains records of all parameter configurations of nodes in the network and serves as the sole decision-maker for all nodes. While this approach is asymptotically optimal, the number of super arms is \((|\mathcal{SF}|\cdot|\mathcal{BW}|\cdot|\mathcal{CF}|\cdot|\mathcal{BW}|)^{N}\), which grows exponentially with the number of nodes. Consequently, this leads to extremely slow convergence and a significantly large regret upper bound \cite{MAMAB}. Therefore, D-LoRa adopts a distributed CMAB approach to avoid the curse of dimensionality.

\begin{figure}[tbp]
\centering
\includegraphics[width=0.9\linewidth]{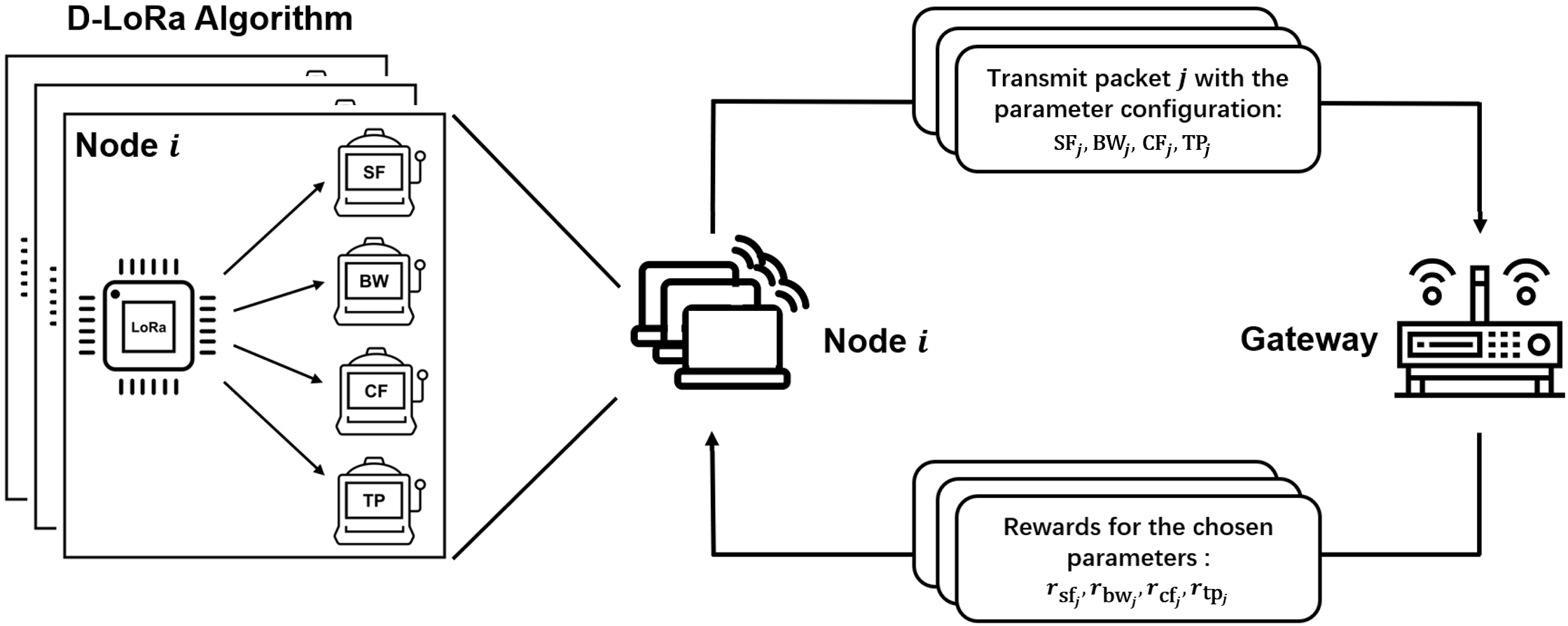}
\caption{Framework of the D-LoRa algorithm}
\label{Framework}
\end{figure}

The framework of D-LoRa is shown in Fig. \ref{Framework}. In D-LoRa, an agent is deployed on each node to determine the parameter configuration for each packet. Specifically, the agent selects a super arm composed of four base arms, corresponding to four parameters, during each decision-making instance, which is the parameter configuration for the subsequent packet transmission. After the transmission, each base arm within the super arm receives the corresponding reward ($r_{\text{sf}}$, $r_{\text{bw}}$, $r_{\text{cf}}$, $r_{\text{tp}}$) for updating its expected reward. Compared to directly selecting a super-arm and updating its expected reward, the action space for each agent in D-LoRa is reduced to $(|\mathcal{SF}|+|\mathcal{BW}|+|\mathcal{CF}|+|\mathcal{BW}|)$.
Furthermore, the distributed training approach facilitates rapid convergence of the algorithm. The update of expected rewards can be expressed as follows:
\begin{equation}
   \overline{R}_{n}^t(a)=\overline{R}_{n}^{t-1}(a)+\frac{1}{T_{n}^t(a)}[r^t_{n}(a)-\overline{R}_{n}^{t-1}(a)],
    \label{Qupdate}
\end{equation}
where $r^t_{n}(a)$ is the observed reward of node $n$ with the base arm $a$ at $t$ th packet transmission, $T_{n}^t(a)$ is the number of times $n$ chooses $a$ until $t$ th packet transmission, and $\overline{R}_{n}(a)$ is $n$'s expected reward of $a$.

UCB1 algorithm is applied to achieve a good balance between the exploration and exploitation in D-LoRa. The estimated reward is defined as:
\begin{equation}
   \widehat{R}_{n}^t(a)=\overline{R}_{n}^t(a)+c\cdot\sqrt{\frac{\log (t)}{2T_{n}^t(a)}},
    \label{UCB}
\end{equation}
where $c\cdot\sqrt{\frac{\log (t)}{2T_{n}^t(a)}}$ is the exploration term and $c$ is the weight factor, which is used to adjust the tendency towards exploration and exploitation. The estimated reward of a super arm is modeled as a linear combination of the estimated rewards of its base arms and the agent selects the super arm with the highest estimated reward at each decision step:
\begin{equation}
   A=\underset{A\in\mathcal{A}}{\text{argmax}}\sum_{a\in A}\widehat{R}_{n}^t(a),
    \label{superarm}
\end{equation}
where $A$ is the super arm consisting of four base arms and $\mathcal{A}$ is the set of super arms.
\subsection{Rewards design}
The rewards design for the base arms in D-LoRa takes the nonlinear trade-off relationships among the performance metrics into consideration to meet the requirements of various application scenarios. We propose metric factors ($i.e., \xi, \zeta, \eta$) to tailor each node’s parameter preference.

\textit{1) SF:} To reduce co-SF collisions and improve the packet transmission rate, nodes should tend to select a smaller SF, provided that packet collisions are avoided. Therefore, $r_{sf}$ is defined as:
\begin{equation}
   r_{\text{sf}_{i}}=\mathbb{I}\{C_i=0\land S_i=0\}+\xi\cdot\frac{\frac{\text{SF}_{i}}{2^{\text{SF}_{i}}}}{ {\textstyle \sum_{\text{SF}_{j}\in\mathcal{SF}}} \frac{\text{SF}_{j}}{2^{\text{SF}_{j}}}},
    \label{term1}
\end{equation}
where $\mathbb{I}\{C_i=0\land S_j=0\}$ is the indicator operator, which is 1 if $p_{i}$ is successfully received.

\textit{2) BW:} Since BW affects RS and ToA, nodes should tend to select a larger BW, provided that the packet can be successfully delivered to the gateway. Therefore, $r_{bw}$ is defined as:
\begin{equation}
   r_{\text{bw}_{i}}=\mathbb{I}\{C_i=0\land S_i=0\}+\zeta\cdot\frac{\text{BW}_{i}}{{\textstyle \sum_{\text{BW}_{j}\in\mathcal{BW}}\text{BW}_{j}}},
    \label{term2}
\end{equation}

\textit{3) CF:} To mitigate co-CF collisions, nodes should select distinct CFs. As CF does not affect the EE and TH, the reward function for the CF excludes the metric factor:
\begin{equation}
   r_{\text{tp}_{i}}=\mathbb{I}\{C_i=0\land S_i=0\},
    \label{term3}
\end{equation}

\textit{4) TP:} To reduce abundant energy consumption, TP should be set to the minimum feasible level while ensuring successful packet reception. $r_{tp}$ is defined as:
\begin{equation}
   r_{\text{tp}_{i}}=\mathbb{I}\{C_i=0\land S_i=0\}+\eta\cdot(1-\frac{\text{TP}_{i}}{{{\textstyle \sum_{\text{TP}_{j}\in\mathcal{TP}}\text{TP}_{j}}}}),
    \label{term4}
\end{equation}

Metric factors can be modified to meet specific performance requirements. In smart agriculture, nodes are battery-powered and require long-term deployment in outdoor environments. This scenario imposes stringent requirements on EE while placing relatively lower demands on PDR and TH. To enhance EE, \(\eta\) can be increased to incentivize nodes to select a lower TP.  In contrast, in industrial monitoring, nodes are powered by external sources and require low-latency transmission of large volumes of data. In this scenario, TH is the primary concern, whereas EE and PDR are of lower priority. By increasing \(\xi\) and \(\zeta\), nodes can be encouraged to select smaller SFs and larger BWs to improve TH.
\subsection{Complexity, Regret and Overhead Analysis}
\textit{Time Complexity:} At each transmission slot, an agent independently selects a base arm for each of the four parameters using the UCB1 policy. Let $K_{\text{total}}=|\mathcal{SF}|+|\mathcal{BW}|+|\mathcal{CF}|+|\mathcal{BW}|$ denote the total number of base arms. For each parameter, the agent computes the estimated reward (UCB1 value) for all its candidate base arms and selects the one with the highest value. The time complexity per decision is linear in the total number of base arms, which is $O\left(K_{\text{total}}\right)$. After each transmission, the agent updates the expected reward and the number of times for the selected base arms, which requires $O\left(1\right)$ operations per parameter.

\textit{Space Complexity:} Each agent $n$ maintains two variables for each base arm: the expected reward $\overline{R}_n^t(a)$ and the number of times $T_n^t(a)$. Thus, the space complexity is $O\left(K_{\text{total}}\right)$, which is linear in the total number of base arms.

\textit{Regret Upper Bound:}  For a given base arm 
$a$ of the agent $n$, its expected reward is $\overline{R}_{n}(a)$ and denote the optimal reward (for the corresponding parameter) by $R_{n}^*(a)$. Define the reward gap as:
\begin{equation}
   \Delta_a=R_n^*(a)-\overline{R}_n(a),
    \label{gap}
\end{equation}
where $\Delta_a>0$ for suboptimal arms. According to \cite{regret}, the expected number of times a suboptimal arm $a$ is selected over $t$ rounds is bounded by:
\begin{equation}
    \mathbb{E}[T^t_n(a)]\leq\frac{8\ln t}{\Delta_a^2}+1+\frac{\pi^2}{3}.
    \label{bound1}
\end{equation}
Therefore, the expected regret from base arm $a$ is:
\begin{equation}
    J^{t}_{\text{regret}}(a)=\Delta_i\mathbb{E}[T^t_n(a)]=O\left(\frac{\ln t}{\Delta_a}\right).
    \label{bound2}
\end{equation}
The reward of a super arm is the sum of the rewards of its four constituent base arms, which are chosen independently. The cumulative regret of the super arm (i.e., the loss due to selecting the suboptimal combination) is the sum of the regrets of each base arms. Therefore, if the minimum gap among all suboptimal base arms is denoted by:
\begin{equation}
    \Delta_{\min}=\min_{a:\Delta_a>0}\Delta_a,
\end{equation}
the overall cumulative regret $J_{\text{regret}}(t)$ of the super arm satisfies:
\begin{equation}
    J^t_{\text{regret}}=O\left(\sum_{i=a}^{K_{\mathrm{total}}}\frac{\ln t}{\Delta_{a}}\right)=O\left(\frac{K_{\mathrm{total}}\ln t}{\Delta_{\mathrm{min}}}\right).
\end{equation}
Since the cumulative regret $J^t_{regret}$ grows as $O(\ln t)$, the average regret per round is:
\begin{equation}
    \frac{J^t_{\text{regret}}}{t}=O\left(\frac{K_{\mathrm{total}}\ln t}{\Delta_{\mathrm{min}}t}\right).
\end{equation}
As $T\to\infty$, the average regret tends to zero:
\begin{equation}
    \lim_{t\to\infty}\frac{J^t_{\text{regret}}}{t}=0,
\end{equation}
which implies that the D-LoRa algorithm is asymptotically optimal since the per-round (or average) regret vanishes in the long run. 

\textit{Qualitative Overhead:} Each node operates independently without inter-node coordination in D-LoRa, which eliminates communication overhead and makes D-LoRa suitable for large-scale LoRa networks. The computation and parameter selection require lightweight arithmetic operations, which are feasible for resource-constrained LoRa nodes. In addition, storing expected rewards and the number of times for the \(K_{\text{total}}\) base arms imposes minimal memory demands. It is worth mentioning that D-LoRa requires a learning phase compared to traditional methods like ADR, which may cause additional overhead. Nevertheless, despite the extra training overhead, D-LoRa demonstrates superior performance in PDR and EE compared to ADR as shown in Section \ref{performance}.
\subsection{Training for Agents}
As shown in Alg. \ref{alg:Train}, each agent has four Expected Reward Lists and Number of Times Lists. We regard a complete transmission process of the LoRa network as an episode. The changes of PDR, EE, and TH in each episode can intuitively show the training effect of D-LoRa. The network is initialized at the beginning of each episode (line 4). When a node needs to transmit a packet, it adapts the LoRa parameters for packet transmission and updates its Lists (lines 7-9). 
\begin{algorithm}[!h]
    \caption{D-LoRa algorithm}
    \label{alg:Train}
    \renewcommand{\algorithmicrequire}{\textbf{Input:}}
    \renewcommand{\algorithmicensure}{\textbf{Output:}}
    \begin{algorithmic}[1]
        \REQUIRE The set of nodes $\mathcal{N}$, Gateway GW, Default LoRa network configuration
        \ENSURE PDR, EE, TH
        \STATE  Initialize the four Expected Reward Lists: $R^\text{SF}_{n_i}$, $R^\text{BW}_{n_i}$, $R^\text{CF}_{n_i}$, $R^\text{TP}_{n_i}$, Number of Times Lists: $T^\text{SF}_{n_i}$, $T^\text{BW}_{n_i}$, $T^\text{CF}_{n_i}$, $T^\text{TP}_{n_i}$ and Total Count $t_{n_i}$ for each $n_{i}\in\mathcal{N}$ 
        \STATE Each node chooses each base arm at least once and update the expected reward and number of selections
        \FOR{$episode = 1$ to $M$}
            \STATE Initialize $\mathcal{P}^{\text{gr}}$ and $\mathcal{P}^{\text{ns}}_{i}$, $\mathcal{P}^{\text{nr}}_{i}$, $\mathcal{P}^{\text{nl}}_{i}$ for each $n_{i}\in\mathcal{N}$
            \STATE Start packet transmission
            \WHILE{Network does not satisfy the stop condition}
                \STATE Each node choose parameters base on Eq. (\ref{UCB})-(\ref{superarm}) for each packet transmission
                \STATE Observe base arm rewards: $r_{\text{sf}}$, $r_{\text{bw}}$, $r_{\text{cf}}$, $r_{\text{tp}}$
                \STATE Update the expected rewards and number of selections of the based arms based on Eq. (\ref{Qupdate})
            \ENDWHILE
        \ENDFOR
        \STATE Output PDR, EE, TH of the last episode
    \end{algorithmic}
\end{algorithm}

\section{Scheme Performance}
\label{sec:exp}
In this section, we analyze the performance of the proposed scheme and four baseline algorithms. At the same time, we derived variant algorithms from D-LoRa towards different network performance metrics.
\subsection{Experiment Setting}
The simulations are carried out on the LoRaWAN simulator LoRaSimPlus\cite{Simulator}. 
We consider a circular LoRa network with the topological radius varying from 1000 m to 2500 m, in which 50 nodes are randomly distributed in the network area as shown in Fig. \ref{topo}. During the simulation, each node continuously sends packets with the same payload size of $20$ bytes to GW. The available LoRa parameters for each node are set as: $\mathcal{SF}=\{7,8,9,10,11,12\}$, $\mathcal{BW}=\{125,250,500\}$ kHz, $\mathcal{CF}=\{470.1,470.3,470.5,470.7,470.9,471.1,471.3,471.5\}$ MHz, $\mathcal{TP}=\{2,4,6,8,10,12,14\}$ dBm. The packet sending interval of the node follows an exponential distribution with the parameter $\lambda=0.25$, which means the average packet sending period of the node is $4$ s. The parameters of the path loss model are set to $\overline{L^{\text{pl}}}(d_{0})=128.95$ dB, $d_{0}=1000$ m, $\gamma=2.32$, $\delta_{1}=7.8$. AWGN follows the zero-mean normal distribution with the standard deviation $\delta_{2}=1$. The weight factor $c$ is set to $2$.
\subsection{Baseline Algorithms}
The four baseline algorithms are shown as follows:
\begin{itemize}
\item \textbf{Random}: Each node configs LoRa parameters randomly for packet transmission.
\item \textbf{Round-Robin}: Each node assigns SF and CF to its packets based on its identity in a round-robin fashion to ensure unique combinations. BW and TP are randomized.
\item \textbf{ADR}\cite{ADR}: ADR selects the minimum available SF and BW combination based on the link budget that the node used to send packets. Meanwhile, the smallest possible TP is assigned to reduce energy consumption.
\item \textbf{RS-LoRa}\cite{RS-LoRa}: RS-LoRa maintains relatively consistent collision probabilities across different SFs. Nodes tend to choose smaller SFs with lower collision probability.
\end{itemize}
\begin{table*}[tp]  
\vspace*{5mm}
  \begin{center}
    \renewcommand{\arraystretch}{1.2}
    \captionsetup{position=above}  
    \caption{Experimental results of the four D-LoRa variants}
    \label{tab:table1}
    \begin{adjustbox}{width=\textwidth}  
        \begin{tabular}{cc|c c c|c c c|c c c|c c c}
          \hline
              \multicolumn{2}{c|}{\multirow{3}{*}{\textbf{Topology Radius(m)}}}&\multicolumn{3}{c|}{\textbf{D-LoRa}}&\multicolumn{3}{c|}{\textbf{D-LoRa-PDR}}&\multicolumn{3}{c|}{\textbf{D-LoRa-EE}}&\multicolumn{3}{c}{\textbf{D-LoRa-TH}}\\
              \cline{3-14}
              \multicolumn{2}{c|}{}& 
              \textbf{PDR}&\textbf{EE}&\textbf{Throughput}& 
              \textbf{PDR}&\textbf{EE}&\textbf{Throughput}& 
              \textbf{PDR}&\textbf{EE}&\textbf{Throughput}& 
              \textbf{PDR}&\textbf{EE}&\textbf{Throughput}\\
               \multicolumn{2}{c|}{} &\textbf{(\%)}& \textbf{(bits/mJ)} & \textbf{(bps)} &\textbf{(\%)} & \textbf{(bits/mJ)} & \textbf{(bps)} &\textbf{(\%)} & \textbf{(bits/mJ)} & \textbf{(bps)} &\textbf{(\%)} & \textbf{(bits/mJ)} & \textbf{(bps)} \\
          \hline
          \multicolumn{2}{c|}{\textbf{1000}} & 90.91&84.22 &573 &95.30 &25.67 &617 &84.14 &125.19 &412 &89.91 &36.69 & 888\\
          \multicolumn{2}{c|}{\textbf{1500}} & 89.83&39.60 &551 &92.14 &23.47 &553 &80.89 &50.79 &357 &83.68 &26.08 &652\\
          \multicolumn{2}{c|}{\textbf{2000}} & 88.30&22.33 &491 &89.46 &21.33 &536 &78.73 &37.69 &381 &83.68 &17.47 &428\\
          \multicolumn{2}{c|}{\textbf{2500}} & 85.81&21.05 &462 &86.70 &17.25 &432 &79.07 &23.15 &348 &77.65 &8.86 &221\\
          \hline
        \end{tabular}
    \end{adjustbox}
  \end{center}
\end{table*}

We derived four variant algorithms of D-LoRa by modifying the metric factors towards different metrics: D-LoRa ($\xi=0, \zeta=0, \eta=1.8$), D-LoRa-PDR ($\xi=0, \zeta=0, \eta=0$), D-LoRa-EE ($\xi=0, \zeta=0, \eta=3.5$), D-LoRa-TH ($\xi=10, \zeta=10, \eta=0$). Due to space limitations in this conference paper, parameter tuning experiments are not presented. For detailed experimental results, please refer to \cite{Simulator}.
\subsection{Performance Comparison}
\label{performance}
\begin{figure}[ht] \label{fig7} 
  \centering
  \begin{minipage}[b]{0.48\linewidth}  
    \centering
    \includegraphics[width=\linewidth]{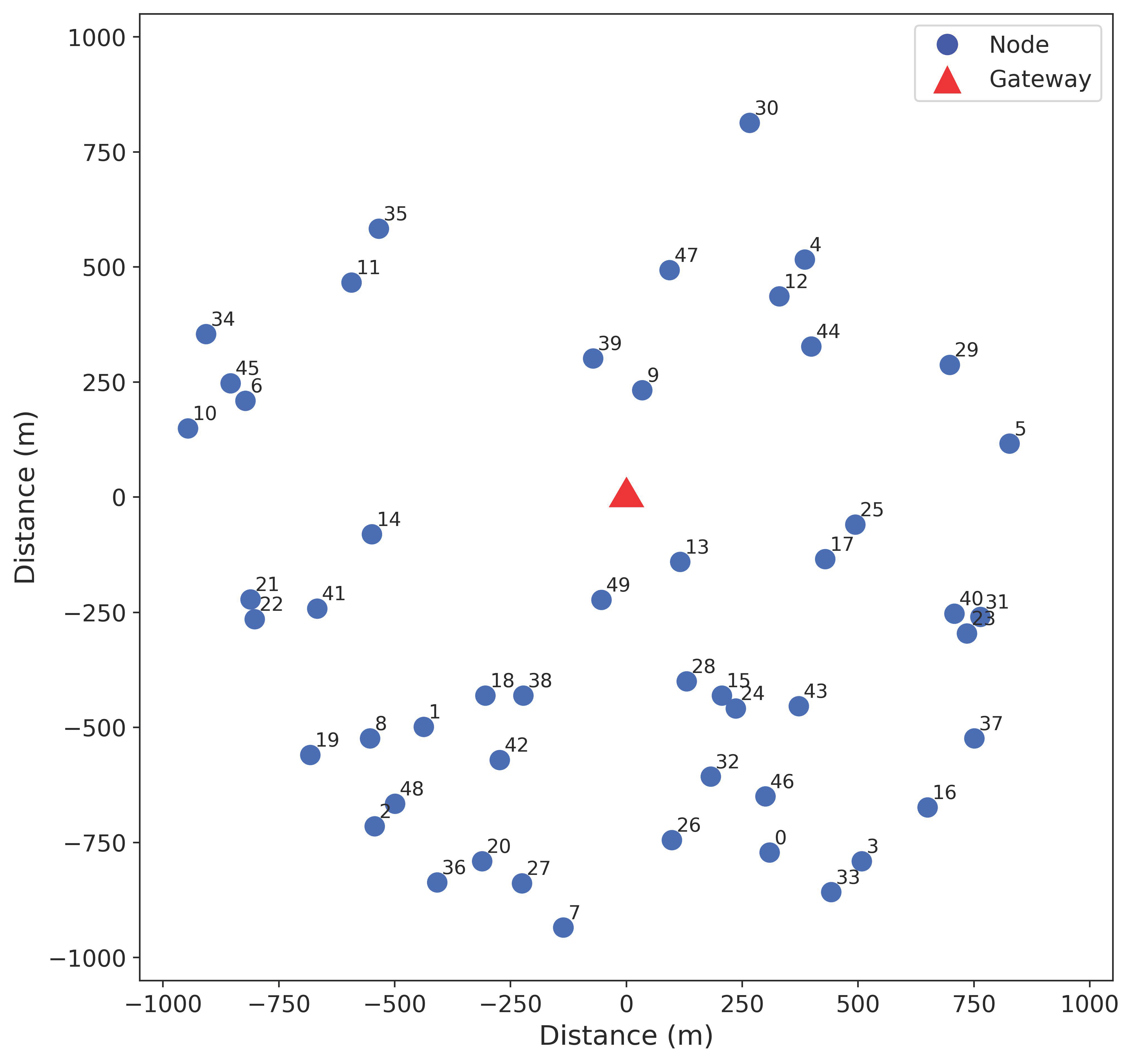} 
    \caption{Environment} \label{topo}
  \end{minipage} 
  \begin{minipage}[b]{0.48\linewidth} 
    \centering
    \includegraphics[width=\linewidth]{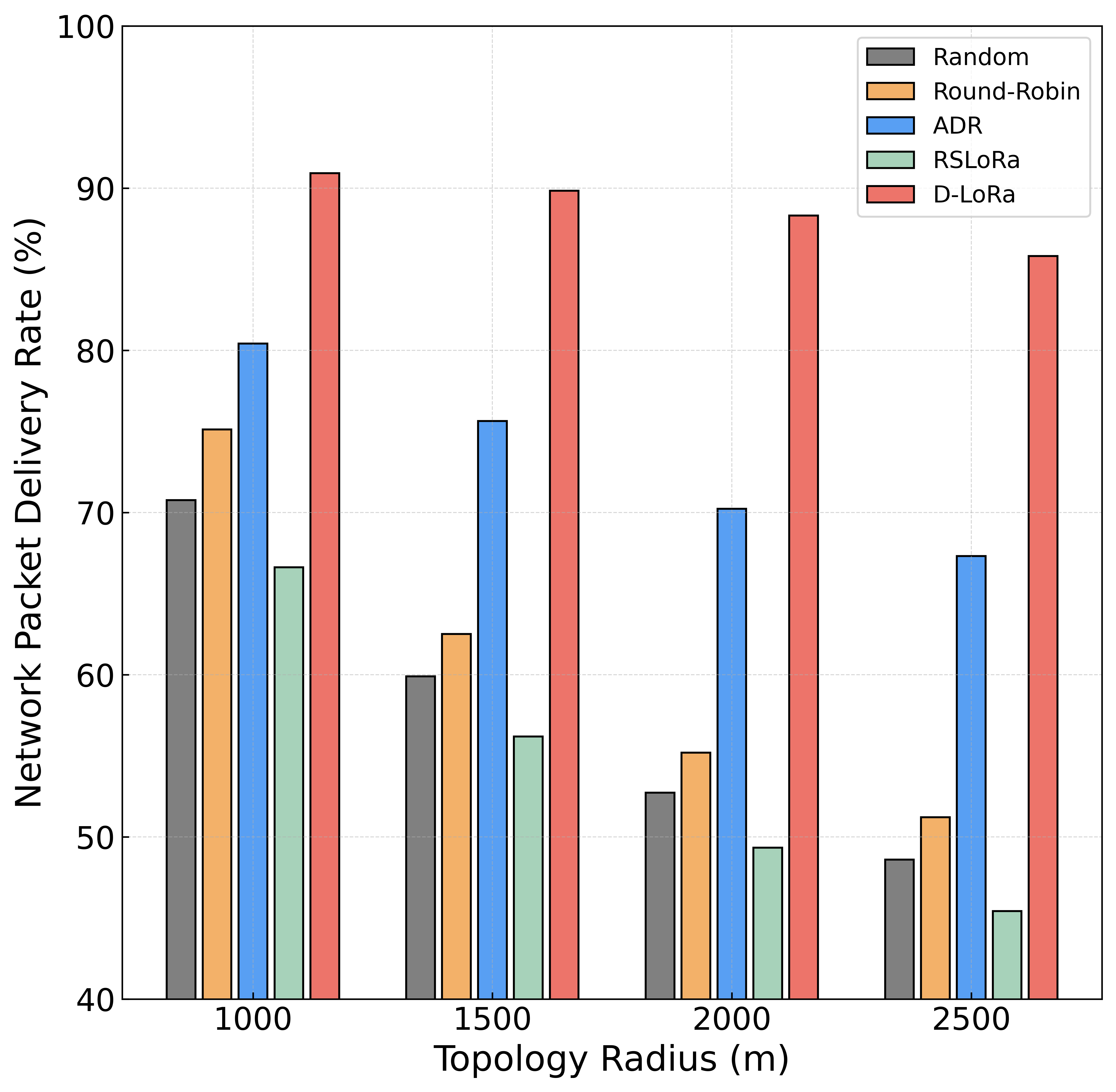} 
    \caption{Network PDR} \label{PDR}
  \end{minipage} 
  \vspace{2mm}  
  \begin{minipage}[b]{0.48\linewidth}  
    \centering
    \includegraphics[width=\linewidth]{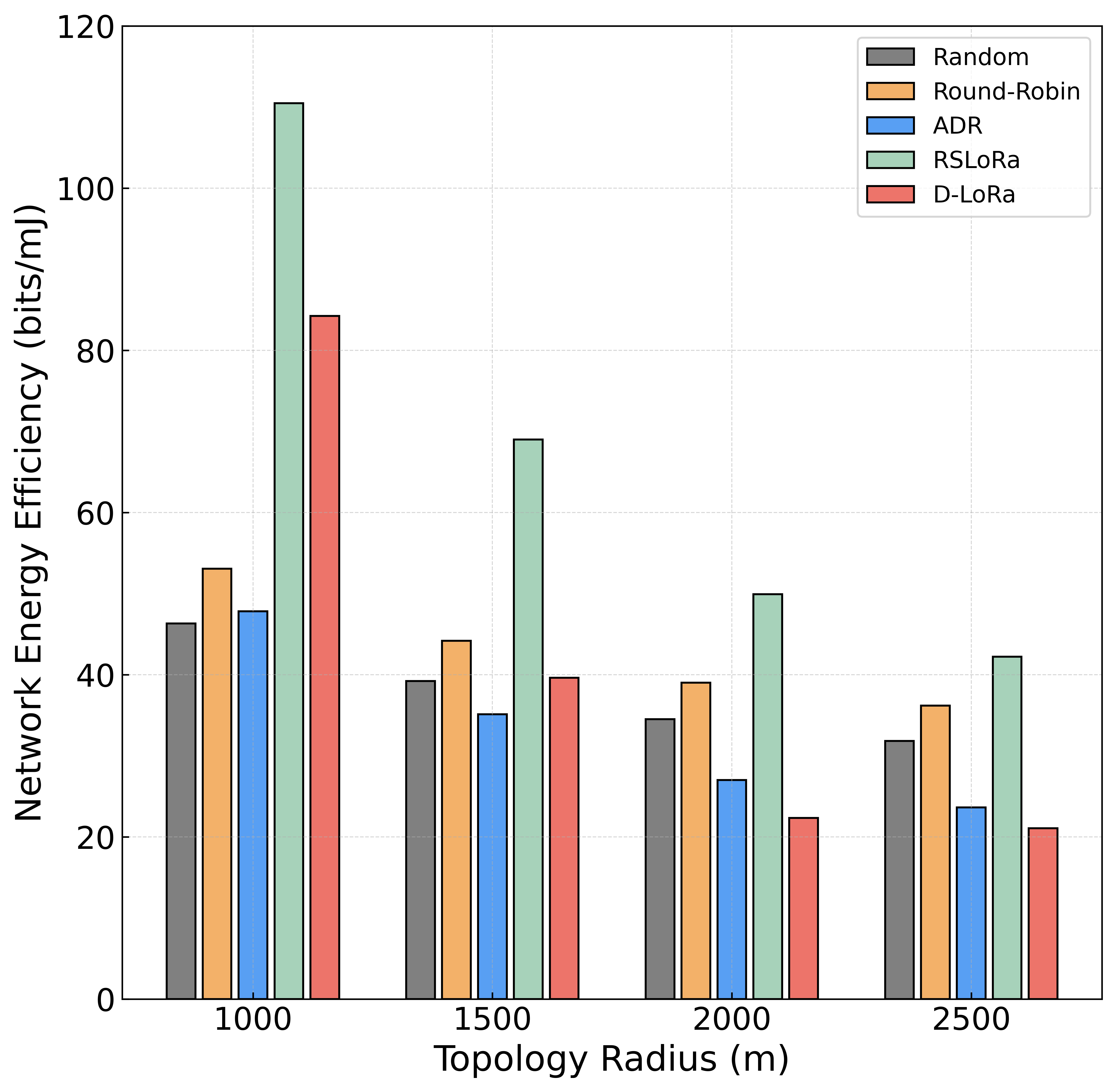} 
    \caption{Network EE}\label{EE}
  \end{minipage}
  \begin{minipage}[b]{0.48\linewidth}  
    \centering
    \includegraphics[width=\linewidth]{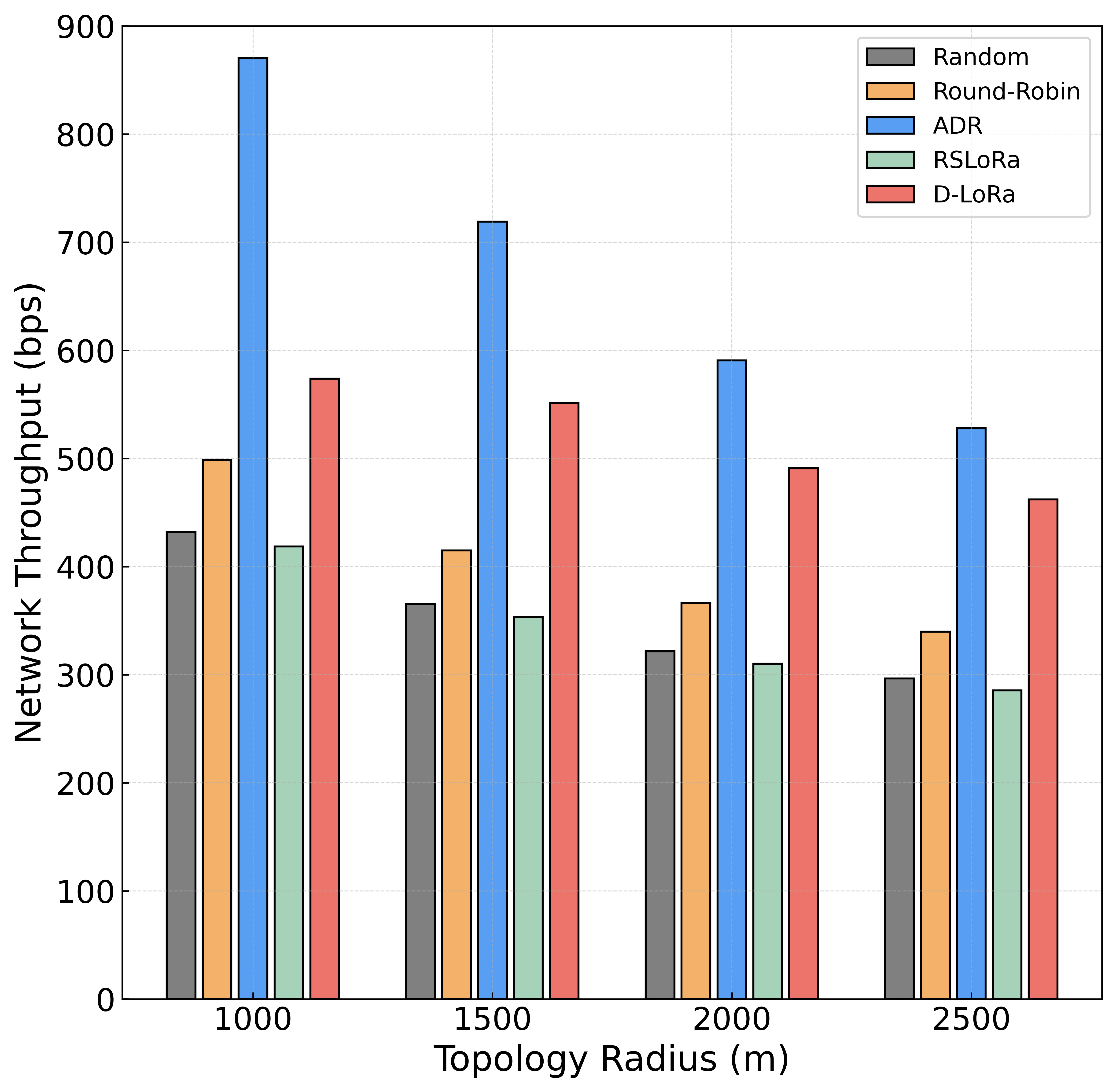} 
    \caption{Network TH}\label{TH}
  \end{minipage} 
\end{figure}
Fig. \ref{PDR}-\ref{TH} illustrates the simulation results of PDR, EE, and TH for D-LoRa and baselines under different radius. Fig. \ref{PDR} demonstrates that D-LoRa significantly improves the PDR, achieving an increase of 10.50\% (at 1000 m) to 18.50\% (at 2500 m) compared to the best-performing baseline algorithm. Furthermore, as the radius increases, the PDR of D-LoRa does not exhibit a notable decline as observed in other algorithms. This superior performance can be attributed to D-LoRa's parameter adaptation mechanism, which leverages historical packet transmission information rather than solely relying on node-gateway distance as in the conventional ADR algorithm.

As shown in Fig. \ref{EE}, RS-LoRa maximizes EE by choosing smaller SFs, reducing transmission time and energy consumption. However, this compromises reliable transmission, which results in its low PDR in Fig. \ref{PDR}. Conversely, D-LoRa prioritizes reliability via adaptive parameter adjustments (large SF/TP and small BW for distant nodes), achieving optimal PDR but higher energy consumption. This trade-off between EE and PDR stems from the dual impact of SF: larger SFs improve reliability (PDR) at the expense of prolonged transmissions (EE). In small radii (e.g., 1000m), D-LoRa approaches RS-LoRa’s EE by favoring smaller SF/TP, demonstrating its adaptability for reliability-centric scenarios. 

As shown in Fig. \ref{TH}, ADR exhibits the highest TH by dynamically optimizing transmission parameters (SF/BW). D-LoRa achieves suboptimal TH, balancing PDR and TH via adaptive parameter configuration for distant nodes. While ADR’s distance-dependent strategy limits flexibility (nodes at similar distances from the gateway choose the same configuration), D-LoRa leverages historical transmission data to optimize reliability, achieving the highest PDR and near-optimal EE in small-scale networks (Fig. \ref{PDR}, \ref{EE}). 

Table \ref{tab:table1} shows that the D-LoRa variants demonstrate different performance trade-offs adapted to different application scenarios. D-LoRa-PDR prioritizes reliability, achieving the highest PDR (e.g., 95.30\% at 1000m) by choosing larger SF and TP, but incurs significant energy inefficiency (EE: 25.67 bits/mJ) and moderate throughput (617 bps). D-LoRa-EE maximizes EE (125.19 bits/mL at 1000m) through minimal SF/TP configurations, yet sacrifices PDR (84.14\%) and TH (412 bps). D-LoRa-TH optimizes TH (888 bps at 1000m) via high-rate transmissions but exhibits rapid performance degradation at larger radii (221 bps at 2500m) and reduced PDR (89.91\%) and EE (36.69 bits/mJ). These results highlight D-LoRa’s adaptability: D-LoRa-PDR suits reliability-critical scenarios (e.g., health monitoring), D-LoRa-EE aligns with energy-constrained deployments (e.g., smart agriculture), and D-LoRa-TH addresses high-density data collection (e.g., industrial monitoring). The ability of D-LoRa to handle the trade-off among performance metrics underscores its versatility in meeting demands of diverse IoT application.

\section{conclusion}
\label{sec:cons}
In this study, we have investigated the network performance optimization problem in a LoRa network. A distributed parameter adaptation algorithm, D-LoRa, has been proposed to solve the complex combinatorial optimization problem. Reward functions tailored for different network performance metrics were designed. Experimental results have validated the superiority of D-LoRa over baseline algorithms.
The flexibility and optimization results of D-LoRa suggest promising applications within LoRa systems. Future research content could involve quantitative overhead analysis of D-LoRa and validating its performance in real-world deployments.

\end{document}